\documentclass{article}

\usepackage{amsmath}
\usepackage{amssymb}
\usepackage{graphicx}
\usepackage{color}

\begin{document}

\title{Noise-assisted Multibit Storage Device}

\author{G. Bellomo$^{(1)}$ \and G. A. Patterson$^{(1)}$ \and P. I. Fierens$^{(2,3)}$ \and D. F. Grosz$^{(1,2,3)}$\\
\small (1) Departamento de F\'isica, FCEN, Universidad de Buenos Aires, Argentina\\
\small (2) Instituto Tecnol\'ogico de Buenos Aires, Argentina\\
\small (3) Consejo Nacional de Investigaciones Cient\'ificas y T\'ecnicas, Argentina
}

\maketitle

\begin{abstract}
In this paper we extend our investigations on noise-assisted storage devices through the experimental study of a loop composed of a single Schmitt trigger and an element that introduces a finite delay. We show that such a system allows the storage of several bits and does so more efficiently for an intermediate range of noise intensities. Finally, we study the probability of erroneous information retrieval as a function of elapsed time and show a way for predicting device performance independently of the number of stored bits.
\end{abstract}

\section{Introduction}
\label{sec:introduction}

In the last few years there has been an increased effort in the search of alternative technologies for computer memory devices (see, e.g., \cite{Wolf.IEEEProceedings.2010} and references therein). This effort is motivated by the perceived near-future end of the ability of current technologies to provide support for the exponential increase of memory capacities as predicted by Moore's law. In this context, we have worked on the concept of memories which can benefit from (and, indeed, work only in) the presence of noise which is typically found in electronic circuits. In a similar vein, logic gates that work with the help of noise have been suggested in, e.g., Refs. \cite{Murali.PRL.2009,Murali.APL.2009}. Building on the work of Refs. \cite{Carusela.PhysRevE.2001,Carusela.PhysicaD.2002}, where a ring of identical oscillators was shown to be able to sustain a traveling wave with the aid of noise long after the harmonic drive signal had been switched off, we showed that a ring of two bistable oscillators is capable of storing a single bit of information in the presence of noise \cite{Fierens.PhysLettA.2010,Ibanez.EPJB.2010,Patterson.PhysA.2010}. Furthermore, we showed that with the addition of small amounts of white Gaussian noise the system outperformed the deterministic (noiseless) case in terms of the probability of erroneous information retrieval. 

In this paper we extend our investigations on noise-assisted storage devices through the experimental study of a loop composed of a single Schmitt trigger (ST) and an element that introduces a finite delay. We show that such a system allows the storage of several bits and does so more efficiently for an intermediate range of noise intensities. Finally, we study the probability of erroneous information retrieval as a function of elapsed time and describe a way of predicting device performance independently of the number of stored bits. 

The proposed memory device can also be considered as a toy model of a long transmission link with in-line nonlinear elements such as saturable absorbers (\cite{Mamyshev.1998,Gammaitoni.RevModPhys.1998}). An ST approximates the behavior of a saturable-absorber and the delay introduced ad hoc models the natural propagation time of the signal in the transmission line. 

Nonlinear delayed loops have been extensively studied (see, e.g., \cite{Aida.IEEEQuantumE.1992,Losson.Chaos.1993,Mensour.PhysLettA.1995,Mensour.PhysLettA.1998,Morse.PhysLettA.2006,Jin.PhysicaA.2007,Middleton.PhysRevE.2007}). This type of systems usually present complex nonlinear behavior, including self-sustained oscillations and chaotic operation regimes. Memory devices that make use of regimes that show multistable behavior of delayed feedback loops have been proposed. Ref. \cite{Aida.IEEEQuantumE.1992} presents a memory device that stores bits coded as particular oscillation modes of a delayed feedback loop with an electro-optical modulator. Similarly, it is shown in \cite{Mensour.PhysLettA.1995} that binary messages can be stored using controlled unstable periodic orbits of a particular class of delay loop differential equations. However, the memory devices proposed in Refs. \cite{Aida.IEEEQuantumE.1992, Mensour.PhysLettA.1995} are not assisted by noise. Refs. \cite{Mensour.PhysLettA.1998} and \cite{Jin.PhysicaA.2007} study the behavior of a delayed loop with a single threshold device and a bi-stable device, respectively, but focusing on the response of the system to a harmonic driving signal.

The remaining of the paper is organized as follows. The working of the proposed device is described in Section \ref{sec:device_description}, while Section \ref{sec:device_results} presents experimental results. Section \ref{sec:conclusions} closes this work with some final remarks and conclusions.

\section{Memory device}
\label{sec:device_description}

\begin{figure}
\begin{center}
\includegraphics[scale=0.3]{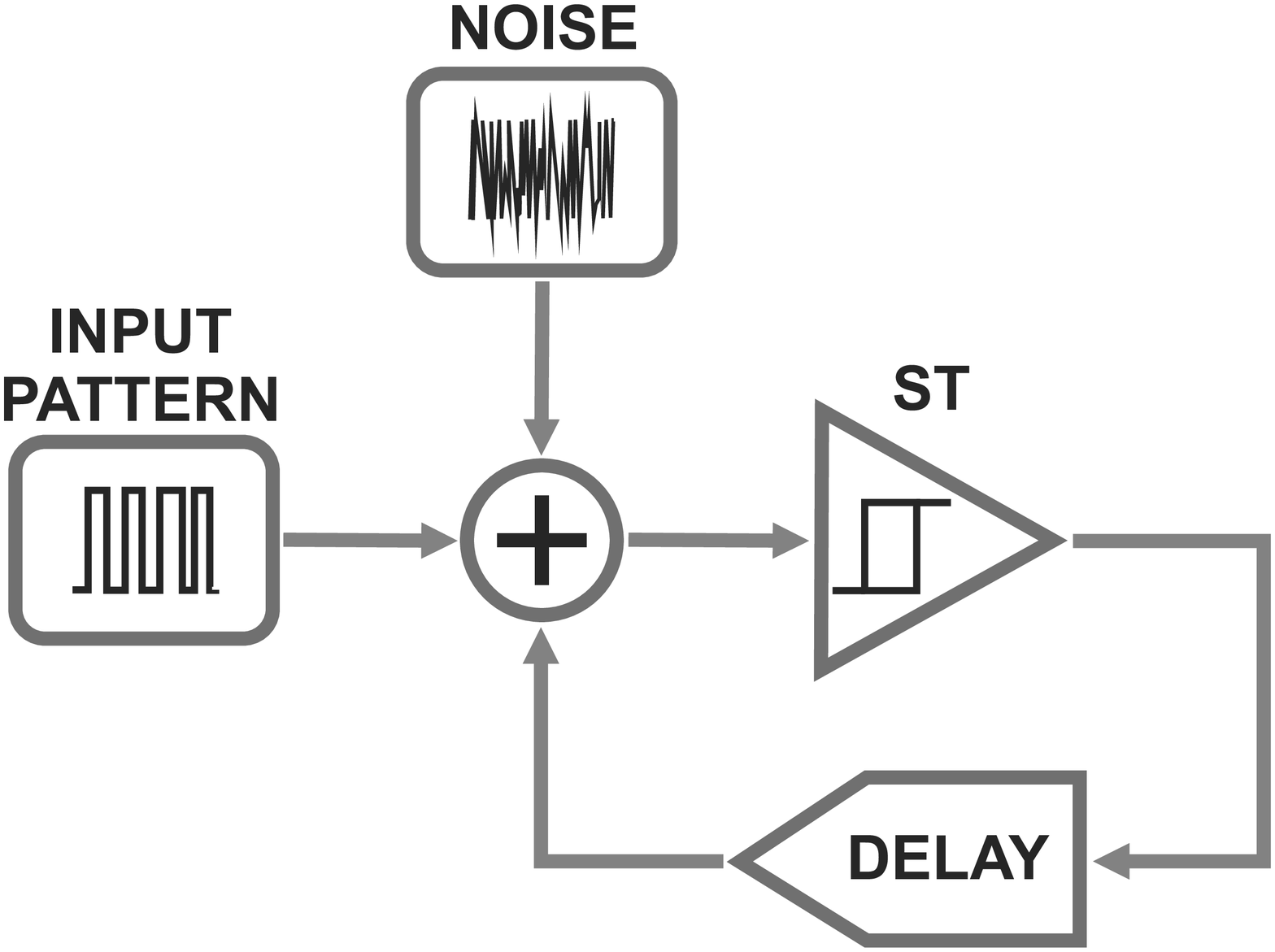}
\end{center}                            
\caption{Schematics of the memory device.}
\label{fig:schematics}
\end{figure}

Fig. \ref{fig:schematics} shows a scheme of the proposed device. One of its main components is a bistable element, a Schmitt trigger in our case. The output of the ST is delayed and then fed back without any other intended alterations besides the addition of (essentially) white Gaussian noise. It should be noted that the output of the Schmitt trigger is subthreshold, that is, it is uncapable of forcing a state change without the assistance of noise. Stored bits are coded as a finite sequence of suprathreshold square pulses. The duration of the sequence is such that it fits entirely in one loop, i.e., it is shorter than or equal to the delay.

In order to retrieve stored bits, the output of the Schmitt trigger must be observed for a time interval equal to the loop delay. In general, a marker is needed to recognize the beginning of a binary message. Since we are not interested in that level of technical detail, in this paper we simply assume that the time elapsed since the bits were stored is perfectly known. Finally, $n$ stored bits are retrieved by comparing the averages of the output of the Schmitt trigger during $n$ consecutive intervals to a fixed threshold. In the experiments, the threshold was varied for each set of fixed parameters in order to obtain the lowest probability of error.

\section{Experimental results}
\label{sec:device_results}

The low and high thresholds of the Schmitt trigger were fixed at $-1$ V and $3$ V, respectively, and the low and high output levels were set at $0$ V and $2$ V, respectively. The input bit pattern was represented by a sequence of pulses with suprathreshold low and high amplitudes of $-8$ V and $10$ V, respectively.

The delay was implemented by means of a in-series pair of 64-bit shift registers clocked with a 100 kHz square wave signal. In this way, the output of the Schmitt trigger is effectively sampled at 100 kHz and its samples are delayed by 1.28 ms. An $n$-bit sequence to be stored occupies the full 1.28 ms-delay, i.e., each bit occupies $1280/n$ $\mu$s. Noise bandwidth was set greater than 100 kHz so that it can be regarded as essentially white where most of the input signal power lies.

\begin{figure}
\begin{center}
\includegraphics[scale=0.3]{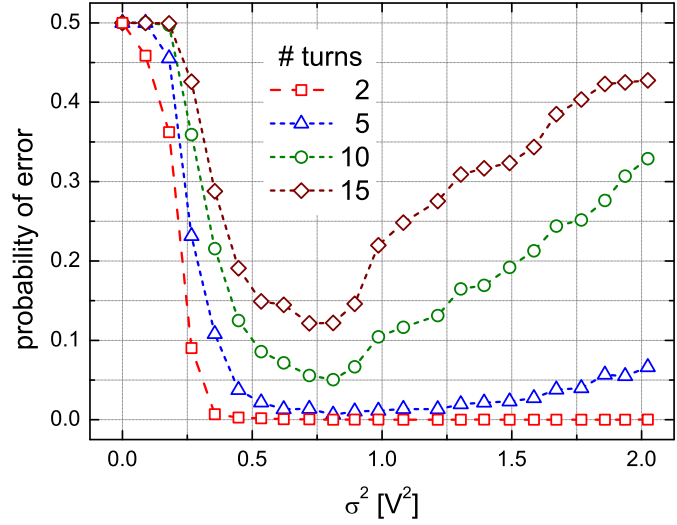}
\end{center}                            
\caption{Probability of error vs. noise intensity when four bits are stored. Different curves correspond to varying elapsed times.}
\label{fig:errorvsnoise}
\end{figure}

Fig. \ref{fig:errorvsnoise} shows the probability of error as a function of the noise intensity for different retrieval times when four bits are stored. Each point corresponds to the average of 1000 realizations and the stored sequence was $0110$. As it is readily observed, there is a range of noise intensities that minimizes the probability of error. Indeed, since the output of the ST is subthreshold, the probability of error is high ($\sim 1/2$) in the absence of noise. The addition of noise helps the subthreshold input signal to drive the Schmitt trigger. Moreover, the probability of a change in the right direction is higher than that in the opposite direction because of the closer proximity to the correct threshold. Therefore, the probability of error initially decreases as the noise intensity increases. However, when the noise intensity is too high with respect to the inter-threshold gap, the probability of error increases because the influence of the driving signal is small with respect to that of noise.

\begin{figure}
\begin{center}
\includegraphics[scale=0.3]{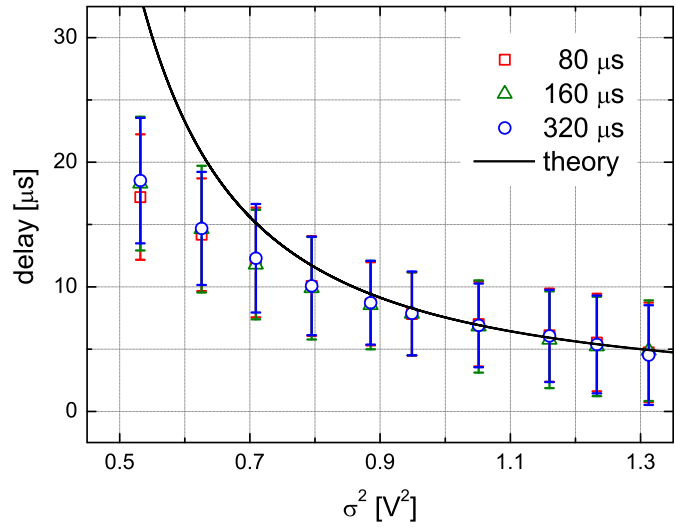}
\end{center}                            
\caption{Mean and standard deviation of the random delay vs. noise intensity. Different curves correspond to different bit lengths. Solid curve: theoretical approximation.}
\label{fig:delayvsnoise}
\end{figure}

Noise at the input of the Schmitt trigger introduces a random delay with a non-negligible mean value. In fact, such delay may be profitably used to build a noise-tuned delay line on the basis of a chain of bistable devices, as it is explained in, e.g., Ref. \cite{Ibanez.FNL.2008}. This random delay needs to be taken into account in order to stay synchronized with the output of the ST, by factoring in its mean value as part of the total loop delay. Fig. \ref{fig:delayvsnoise} shows the mean value and the dispersion of the ST delay as a function of the noise intensity. The random delay was estimated based on the cross-correlation between the input and the output of a Schmitt trigger, where the input consisted of 128 randomly generated pulses of fixed duration. Each point in Fig. \ref{fig:delayvsnoise} is the result of averaging 1000 realizations. It is interesting to note that the stochastic delay decreases as the noise intensity ($\sigma^2$) increases. It can also be observed that there is a range of noise intensities for which the standard deviation of the delay is not too high as compared to its mean value, i.e., in this range timing jitter of the transmitted signal is tolerable. Finally, it must be noted that the delay measured on the basis of the cross-correlation lacks of true meaning in the regions of low and high noise intensities where the output barely resembles the input signal. Fig. \ref{fig:delayvsnoise} also shows a theoretical approximation to the mean delay obtained by modeling noise as an Ornstein-Uhlenbeck (OU) process, with bandwidth $f_c$ equal to the measured bandwidth. Since the output of the ST follows input transitions whenever the instantaneous noise amplitude is able to bridge the gap to the closest threshold ($\Delta V = 1$ V), we approximated the delay by the expected time for the OU process to go from $0$ V (its stationary mean) to $\Delta V$, that is (see, e.g., Ch. 5 of Ref. \cite{Gardiner2004})
\begin{equation}
\left<\tau_{tr}\right> \approx \frac{1}{2\sqrt{\pi}f_c} \int\limits_0^{\frac{\Delta V}{\sqrt{\sigma^2/2}}} e^{x^2}\left[1+\text{erf}(x)\right] dx, 
\label{eq_unique}
\end{equation}
where $\sigma^2$ is the measured noise power. Given that there is essentially no delay when there are no transitions in the driving signal, we estimated the actual mean delay as $\left<\tau\right> \approx \left<\tau_{tr}\right>/2$. 

\begin{figure}
\begin{center}
\includegraphics[scale=0.3]{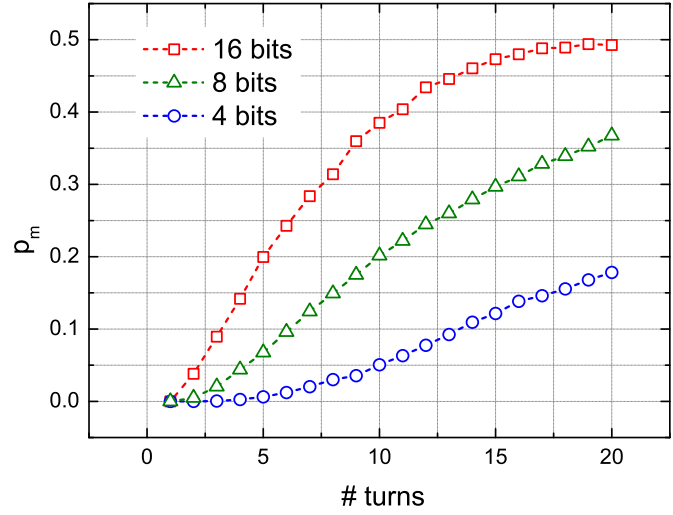}
\end{center}                            
\caption{Minimum probability of error vs. elapsed time.}
\label{fig:minimumerror}
\end{figure}

The experiment was repeated for $8$- and $16$-bit sequences. Fig. \ref{fig:minimumerror} shows the minimum probability of error ($p_m$) in each case as a function of time. As it can be observed, not only the performance for a given number of bits deteriorates with time, but also with an increasing number of stored bits. This  can be understood as follows. The stochastic delay must be short enough so that the ST output is able to follow a transition at its input before the start of the following bit. Since the bit duration decreases as the number of stored bits increases, the noise intensity required to get a shorter random delay increases (Eq. \ref{eq_unique}) with the number of bits (see Fig. \ref{fig:delayvsnoise}). In other words, the optimal noise intensity increases with the number of stored bits. However, a higher noise intensity also implies a higher probability of ST transitions in the wrong direction and, hence, a performance degradation. 

\begin{figure}
\begin{center}
\includegraphics[scale=0.3]{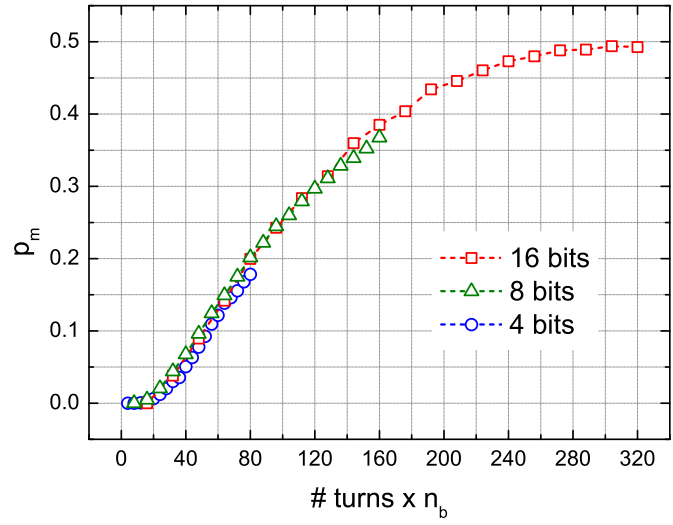}
\end{center}                            
\caption{Minimum probability of error vs. elapsed time normalized to the bit duration. $n_b$: number of stored bits.}
\label{fig:normalizedminimumerror}
\end{figure}

Although Fig. \ref{fig:minimumerror} points to a performance which depends on the number of bits stored, there is a scaling that allows a unified description. Indeed, as Fig. \ref{fig:normalizedminimumerror} shows, the probability of error exhibits the same behavior when time is normalized to the bit duration. 

\section{Conclusions}
\label{sec:conclusions}

In summary, we presented a multibit storage device comprised of a bistable element in a loop configuration which works only with the aid of noise. Moreover, there is a range of noise intensities that minimizes the probability of erroneous information retrieval. In order to measure the probability of error, the random input-output delay at the Schmitt trigger had to be taken into account. We experimentally determined this delay and proposed a theoretical model which agrees very well with measurements. 

We also studied the behavior of the mimimun probability of error as a function of time. We found that performance deteriorates with time and with the number of stored bits. However, we showed that the probability of error is independent of the number of bits when the elapsed time is normalized to the bit duration, a fact of relevance when considering practical implementations of the proposed memory device.

Finally, we believe that this type of storage devices may be of interest as an alternative technology to push forward the validity of Moore's law as current devices have to deal with decreasing noise margins. 

\section*{Acknowledgement}

We gratefully acknowledge financial support from ANPCyT under Project PICTO-ITBA \#31176.

\bibliography{multibit}

\begin{thebibliography}{10}

\bibitem{Aida.IEEEQuantumE.1992}
T.~Aida and P.~Davis.
\newblock Oscillation modes of laser diode pumped hybrid bistable system with
  large delay and application to dynamical memory.
\newblock {\em IEEE Journal of Quantum Electronics}, 28(3):686--699, mar 1992.

\bibitem{Carusela.PhysRevE.2001}
M.~F. Carusela, R.~P.~J. Perazzo, and L.~Romanelli.
\newblock Stochastic resonant memory storage device.
\newblock {\em Phys. Rev. E}, 64:031101, 2001.

\bibitem{Carusela.PhysicaD.2002}
M.~F. Carusela, R.~P.~J. Perazzo, and L.~Romanelli.
\newblock Information transmission and storage sustained by noise.
\newblock {\em Physica D}, pages 177--183, 2002.

\bibitem{Fierens.PhysLettA.2010}
P.~I. Fierens, S.~A. Ib{\'a}{\~n}ez, R.~P.~J. Perazzo, G.~A. Patterson, and
  D.~F. Grosz.
\newblock A memory device sustained by noise.
\newblock {\em Physics Letters A}, 374(22):2207--2209, May 2010.

\bibitem{Gammaitoni.RevModPhys.1998}
L.~Gammaitoni, P.~H{\"a}nggi, P.~Jung, and F.~Marchesoni.
\newblock Stochastic resonance.
\newblock {\em Rev. Mod. Phys.}, 70(1):223--287, Jan. 1998.

\bibitem{Gardiner2004}
C.~W. Gardiner.
\newblock {\em Handbook of Stochastic Methods: for Physics, Chemistry and the
  Natural Sciences}.
\newblock Springer Series in Synergetics. Springer, 3rd edition, April 2004.

\bibitem{Ibanez.FNL.2008}
S.~A. Ib{\'a}{\~n}ez, P.~I. Fierens, R.~P.~J. Perazzo, and D.~F. Grosz.
\newblock Time delay properties of a stochastic-resonance information
  transmission line.
\newblock {\em Fluctuation and Noise Lett.}, 8(3-4):L315--L321, 2008.

\bibitem{Ibanez.EPJB.2010}
S.~A. Ib{\'a}{\~n}ez, P.~I. Fierens, R.~P.~J. Perazzo, G.~A. Patterson, and
  D.~F. Grosz.
\newblock On the dynamics of a single-bit stochastic-resonance memory device.
\newblock {\em Eur. Phys. J. B}, 76:49--55, 2010.

\bibitem{Jin.PhysicaA.2007}
Yanfei Jin and Haiyan Hu.
\newblock Coherence and stochastic resonance in a delayed bistable system.
\newblock {\em Physica A}, 382(2):423--429, 2007.

\bibitem{Losson.Chaos.1993}
J.~{Losson}, M.~C. {Mackey}, and A.~{Longtin}.
\newblock {Solution multistability in first-order nonlinear differential delay
  equations}.
\newblock {\em Chaos}, 3:167--176, April 1993.

\bibitem{Mamyshev.1998}
P.~V. Mamyshev.
\newblock All-optical data regeneration based on self-phase modulation effect.
\newblock In {\em 24th European Conference on Optical Communication,
  ECOC’98}, volume~1, pages 475--476, 1998.

\bibitem{Mensour.PhysLettA.1995}
Boualem Mensour and Andr{\'e} Longtin.
\newblock Controlling chaos to store information in delay-differential
  equations.
\newblock {\em Physics Letters A}, 205(1):18--24, 1995.

\bibitem{Mensour.PhysLettA.1998}
Boualem Mensour and Andr{\'e} Longtin.
\newblock Synchronization of delay-differential equations with application to
  private communication.
\newblock {\em Physics Letters A}, 244(1-3):59--70, 1998.

\bibitem{Middleton.PhysRevE.2007}
J.~W. Middleton, E.~Harvey-Girard, L.~Maler, and A.~Longtin.
\newblock Envelope gating and noise shaping in populations of noisy neurons.
\newblock {\em Phys. Rev. E}, 75(2):021918, Feb 2007.

\bibitem{Morse.PhysLettA.2006}
Robert Morse and Andr{\'e} Longtin.
\newblock Coherence and stochastic resonance in threshold crossing detectors
  with delayed feedback.
\newblock {\em Physics Letters A}, 359(6):640--646, 2006.

\bibitem{Murali.APL.2009}
K.~Murali, I.~Rajamohamed, Sudeshna Sinha, William~L. Ditto, and Adi~R.
  Bulsara.
\newblock Realization of reliable and flexible logic gates using noisy
  nonlinear circuits.
\newblock {\em Applied Physics Lett.}, 95(19):194102, 2009.

\bibitem{Murali.PRL.2009}
K.~Murali, Sudeshna Sinha, William~L. Ditto, and Adi~R. Bulsara.
\newblock Reliable logic circuit elements that exploit nonlinearity in the
  presence of a noise floor.
\newblock {\em Phys. Rev. Lett.}, 102(10):104101, 2009.

\bibitem{Patterson.PhysA.2010}
G.~A. Patterson, A.~F. Goya, P.~I. Fierens, S.~A. Ib{\'a}{\~n}ez, and D.~F.
  Grosz.
\newblock Experimental investigation of noise-assisted information transmission
  and storage via stochastic resonance.
\newblock {\em Physica A}, 389(9):1965--1970, 2010.

\bibitem{Wolf.IEEEProceedings.2010}
S.~A. Wolf, J.~Lu, M.~R. Stan, E.~Chen, and D.~M. Treger.
\newblock The promise of nanomagnetics and spintronics for future logic and
  universal memory.
\newblock {\em Proceedings of the IEEE}, 98(12):2155--2168, Dec. 2010.

\end{thebibliography}
\bibliographystyle{plain}

\end{document}